\documentclass{article}
\usepackage[textwidth=16cm]{geometry}
\usepackage{graphicx}
\usepackage{amssymb,amsmath}
\usepackage{cite}
\DeclareGraphicsExtensions{.eps,.jpg,.pdf,.ps} 
\begin{document}
\bibliographystyle{unsrt}

\def\E{{\bf E}}
\def\H{{\bf H}}
\def\D{{\bf D}}
\def\B{{\bf B}}
\def\J{{\bf J}}
\def\r{{\bf r}}

\def\Acal{{\mathcal{A}}}
\def\Ccal{{\mathcal{C}}}
\def\Cf{{\mathcal{C}'}}
\def\Dcal{{\mathcal{D}}}
\def\Gcal{{\mathcal{G}}}
\def\Aone{{\Acal}_1}
\def\Azero{{\Acal}_0}
\def\Azeroinv{{\Acal}_0^{-1}}
\def\Cone{{\Ccal}_1}
\def\Czero{{\Ccal}_0}
\def\Czeroinv{{\Ccal}_0^{-1}}
\def\Done{{\Dcal}_1}
\def\Dzero{{\Dcal}_0}
\def\Dzeroinv{{\Dcal}_0^{-1}}
\def\Bcal{{\mathcal{B}}}
\def\Bzeroinv{{\Bcal}_0^{-1}}
\def\bem{$}
\def\eem{$}
\def\beq{\begin{equation}}
\def\eeq{\end{equation}}
\def\Zf{{Z_e}}
\def\Zm{{Z_{mod}}}
\def\Rf{{R_e}}
\def\Amean{\overline{\Acal_1}}
\def\Bmean{\overline{\Bcal_1}}
\def\Cmean{\overline{\Ccal_1}}
\def\Cfmean{\overline{\Cf_1}}
\def\Zmean{{\Lambda}} 

\input epsf

\title{{\bf Random scattering by rough surfaces
    with spatially varying impedance}}  


\author{N S Basra$^{1,2}$, M Spivack$^1$, and O Rath Spivack$^1$}

\font\eightrm=cmr8

\maketitle

{\eightrm
\begin{center}
$^1${Department of Applied Mathematics and Theoretical Physics,
                  Centre for Mathematical Sciences,
                      University of Cambridge,
                          Wilberforce Rd,
                  Cambridge CB3 0WA, UK \\
$^2$ Current address:  Upton Court Grammar School, Lascelles Road,
Slough, SL3 7PR, UK 
}
\end{center}
}

\begin{abstract}
A method is given for evaluating electromagnetic scattering by an
irregular surface with spatially-varying impedance.
This uses an operator expansion with respect to impedance variation and allows
examination of its effects and the resulting modification of the field
scattered by the rough surface.  For a fixed rough surface and
randomly varying impedance, expressions 
are derived for the scattered field itself, and for the coherent field
with respect to impedance variation for both flat and rough
surfaces in the form of effective impedance conditions.

%
\end{abstract}

\section{Introduction\label{intro}}

\par

Many applications of wave scattering from rough surfaces are complicated by 
the involvement of further scattering mechanisms
\cite{dragna,ostashev,sarabandi2,hatziioannou,giovannini,guerin}.  
Radar propagating over a sea surface, for example, may 
encounter spatially varying impedance due to surface inhomogeneities
\cite{depine1,brud1,blumberg1}, or
refractive index variations in the evaporation duct
\cite{brown,hatziioannou2}. 
This is an even greater problem in remote sensing 
over forest or urban terrain \cite{li,sarabandi2}.
Roughness is often the dominant feature but
impedance variation may produce further multiple scattering.
The great majority of theoretical and numerical studies 
nevertheless treat such effects in
isolation
\cite{ogilvy,voronovich,watson,dragna,ostashev}. Of particular note are
the elegant studies of admittance variation by
\cite{dragna}, who obtain analytical solutions by applying Bourret
approximation to a Dyson equation, and of impedance variation
by \cite{ostashev} who derive intensity fluctuation statistics.
Experimental validation of scattering models in complex environments
remains a major difficulty, exacerbated by the lack of 
detailed environmental information, and it is therefore crucial
to distinguish and identify sources of scattering.
In addition, while numerical computation in these cases 
may be feasible for the perfectly reflecting surface,
it can become prohibitive for more complex environments,
particularly in seeking statistics from multiple
realisations.

These considerations are the motivation for this paper. 
The main purpose is to provide an efficient means to evaluate
the effect of impedance variation and its interaction with
surface roughness; in addition we derive
descriptions of the resulting coherent or mean field (averaged
with respect to impedance variation) for
an irregular surface.   (For random surfaces the field may averaged further
with respect to the rough surface in special cases, although this will
be tackled more fully in a later paper and is only sketched here.) 
In order to do this an operator expansion is used: Surface currents
from which scattered fields are determined
are expressed as the solution of an integral equation, in which the
effect of impedance variation is separated from the mean impedance. 
The solution is written in terms of the inverse of the governing
integral operator, and provided the impedance variation about its mean
is moderate,  this inversion can be expanded about the leading term.  This is
carried out here for 2-d problems, for a TE incident field.  
For the coherent field this also leads to expressions for equivalent effective
impedance conditions.  

The paper is organised as follows: Governing equations are set out in
section (\ref{equations}) and the operator expansion is given in
(\ref{varying}).
Section (\ref{coherent}) gives mean field with respect to impedance
variation for a fixed rough surface. The procedure for extending to
averages over randomly rough surfaces is briefly outlined.
Some remarks are given in (\ref{conclusions})
regarding the generalisation to TM, and to the fully 3-dimensional case.
The work here is based in part on results originally presented in
\cite{basra}.

\section{Governing equations\label{equations}}

Consider the wavefield above a rough surface with varying impedance in a
2-dimensional medium, with coordinates $(x,z)$ where $x$ is the
horizontal and $z$ the vertical, directed upwards.
The incident electric field $E$ is assumed to be time-harmonic, with 
time-dependence $\exp(-i\omega t)$, say, and
can be taken to be either horizontally (TE) or vertically (TM) plane
polarized. We can suppress the time-dependence and consider the time-reduced
component, and for the moment will restrict attention to an incident TE field.
Denote the surface profile by $\zeta(x)$, 
with impedance $Z=Z_0 + Z_r$
where $Z_0$ is a constant reference value and $Z_r$ is
spatially-varying.

 The variation  $Z_r$ is due to varying (known) material properties
in the adjacent medium or along the boundary.
When ensemble averages are taken it will be
assumed that $Z_r$ is continuous and statistically stationary in $x$, 
with mean zero and scaled variance  $<(Z_r/Z_0)^2> = \sigma_I^2$.
It will also be assumed that $Z_r$ is not large compared with $Z_0$,
in the sense that the root mean square of its
modulus is less than $|Z_0|$. Consequently $\sigma_I ~<~ 1$. This corresponds to a relatively
high-contrast interface.

Where we treat the surface $\zeta(x)$ as being random, we will
assume it has mean zero and is statistically stationary, and we denote
its variance by $\sigma_S^2$ and its autocorrelation function by
$\rho(\xi)$, where $\xi$ is spatial separation.  Thus the mean surface
plane lies in $z=0$. 
We will also assume the surface and impedance functions are independent.

Here and below, single angled brackets $<~\cdot~>$,
or for compactness an overbar, denotes ensemble averages with respect
to impedance variation.  
Ensemble averages with respect to both impedance variation and
randomly rough surface may be denoted by double angled-brackets $\ll
~\cdot~ \gg$.

The field $E$ in the upper medium obeys the Helmholtz 
wave equation $(\nabla^2+k^2)E=0$ where $k$ is the wavenumber.
Denote by $G$ the free space Green's function, so that (in the
2-dimensional case) $G$ is the zero order Hankel function of the first kind,
\beq 
G(\r,\r') = \frac{1}{4i} H_0^{(1)} (k|\r-\r'|) .
\eeq
The total field $E$ along the surface is then given by the solution of 
a Helmholtz integral equation (see also \cite{dragna,desanto}) as follows:
\begin{equation}E_{inc}(\r_s) ~=~
\frac{1}{2}E(\r_s)~-
~\int_{z=\zeta(x)}\left[{\partial G(\r_s,\r')\over\partial
n}+{ik_0G(\r_s,\r')\over
Z_0+Z_r(x')}\right] E(\r')dS'\rm.\label{inteq}\end{equation} 
where $\r_s$ here is an arbitrary surface point $(x,\zeta(x))$, and 
$\r'=(x',\zeta(x'))$. Elsewhere in the upper half space 
the field can be written as a boundary integral: 
\begin{equation}E(\r) ~=~
~\int_{z=\zeta(x)}\left[{\partial G(\r,\r')\over\partial
n}+{ik_0G(\r,\r')\over
Z_0+Z_r(x')}\right] E(\r')dS'\rm.\label{inteq2}\end{equation} 
where now $\r=(x,z)$ represents a general point in the upper medium.
 (The right-hand-side of equation (\ref{inteq}) is an
 operator from functions on the real line to itself, and the same
 holds for (\ref{inteq2}) if $\r$ is, for example, restricted to a
 line at fixed $z$ parallel to $x$.)

\section{Rough surface with varying impedance\label{varying}}

\subsection{General case\label{general}}

We first derive the operator expansion for the general case of an irregular variable impedance 
boundary, and will later deal with special case of a flat variable-impedance boundary. 
This has been studied by many authors in various parameter regimes. 
Analytical treatment for the statistical averages will be discussed in the subsequent section.

We first write 
\begin{equation}{1\over Z_0+Z_r}\equiv{1\over Z_0}-{Z_r\over
Z_0(Z_0+Z_r)}.
\label{7.17}\end{equation} 

For a rough surface $z=\zeta(x)$ with impedance $Z=Z_0+Z_r(x)$
integral equation (\ref{inteq}) then becomes
\begin{equation}
E_{inc}(\r)=({\Ccal}_0+{\Ccal}_1)E(\r),
\label{eq:8.10'}\end{equation} 
where 
\begin{equation}{\Ccal}_0(~\cdot~)={1\over
2}(~\cdot~)~-~\int_{z=\zeta(x)}\left[{\partial G(\r,\r')\over\partial
n}+{ik_0G(\r,\r')\over
Z_0}\right](~\cdot~)dS'\rm. 
\label{eq:8.5'}
\end{equation} 
and $\Ccal_1$ contains the dependence on impedance variation $Z_r$,
\begin{equation}
{\Ccal}_1(~\cdot~)={ik_0\over
Z_0}\int_{z=\zeta(x)}{Z_r(x')G(\r,\r')\over
Z_0+Z_r(x')}(~\cdot~)dS'.
\label{ccal1}
\end{equation} 
Even when the impedance is constant, so that $\Ccal_1=0$,
there is no closed-form analytical solution
and in general for individual realisation ${\Ccal}_0^{-1}E_{inc}(\r)$ 
must be evaluated numerically.

The solution of (\ref{eq:8.10'}) can be written
\begin{equation}
E(\r)=({\Ccal}_0+{\Ccal}_1)^{-1}E_{inc}(\r).
\label{eq:8.11'}\end{equation} 

The inverse can formally be expanded to give
\begin{equation}({\Czero}+{\Cone})^{-1}\equiv{\Czero}^{-1}-
({\Czero}^{-1}{\Cone}){\Czero}^{-1} +
({\Czero}^{-1}{\Cone})^2{\Czero}^{-1}-\ldots    
\label{eq:8.12}\end{equation} 
Since, by assumption, the effect of the term $\Cone$ is not large, the
series can be assumed to converge, and 
the resulting equation may be truncated to obtain an approximation to
the field $E(\r)$ along the surface:
\begin{equation}E(\r)\cong
{\Czeroinv}E_{inc}(\r)~-~
{\Czeroinv}\left[{\Cone} {\Czeroinv}E_{inc}(\r)\right].\label{eq:8.13} 
\end{equation} 

The first term ${\Czeroinv}E_{inc}$ in this expression corresponds to
constant impedance $Z_0$,  
but in general it is non-specular due to the irregular surface. The
second term accounts  
for the diffraction arising from interaction between impedance
variation $Z_r$ and surface profile $\zeta$. Once the first term has
been  
obtained, the remaining term is evaluated by applying $\Ccal_1$ and
solving again 
for $\Ccal_0^{-1}$, with the term in square brackets acting as a new
driving field.   From this, the field away from the surface is
obtained from boundary integral  
(\ref{inteq2}).

This formulation conveniently captures the balance between scattering
mechanisms, and in important cases is efficient for numerical
calculation of the field statistics with respect to impedance 
variation, as well as allowing theoretical estimates of the field statistics to
be obtained. 
For a single realisation of $\zeta(x)$ and $Z(x)$, numerical
evaluation is generally needed.  
Inversion of the integral equation (\ref{eq:8.10'}) is highly costly
computationally. However, in several important regimes including low
grazing angles 
highly efficient methods are available (eg
\cite{Kapp,spivack1,spivack2,brennan}) which cannot be applied directly to the
full integral equation (\ref{eq:8.10'}). In addition equation
(\ref{eq:8.13}) allows analytical treatment in special cases for the
mean field due to a random impedance and either a fixed surface, or a
randomly rough surface.  

\begin{figure}
\par\epsfxsize=16cm
\hskip 1.5 true cm
\includegraphics[height=7cm]{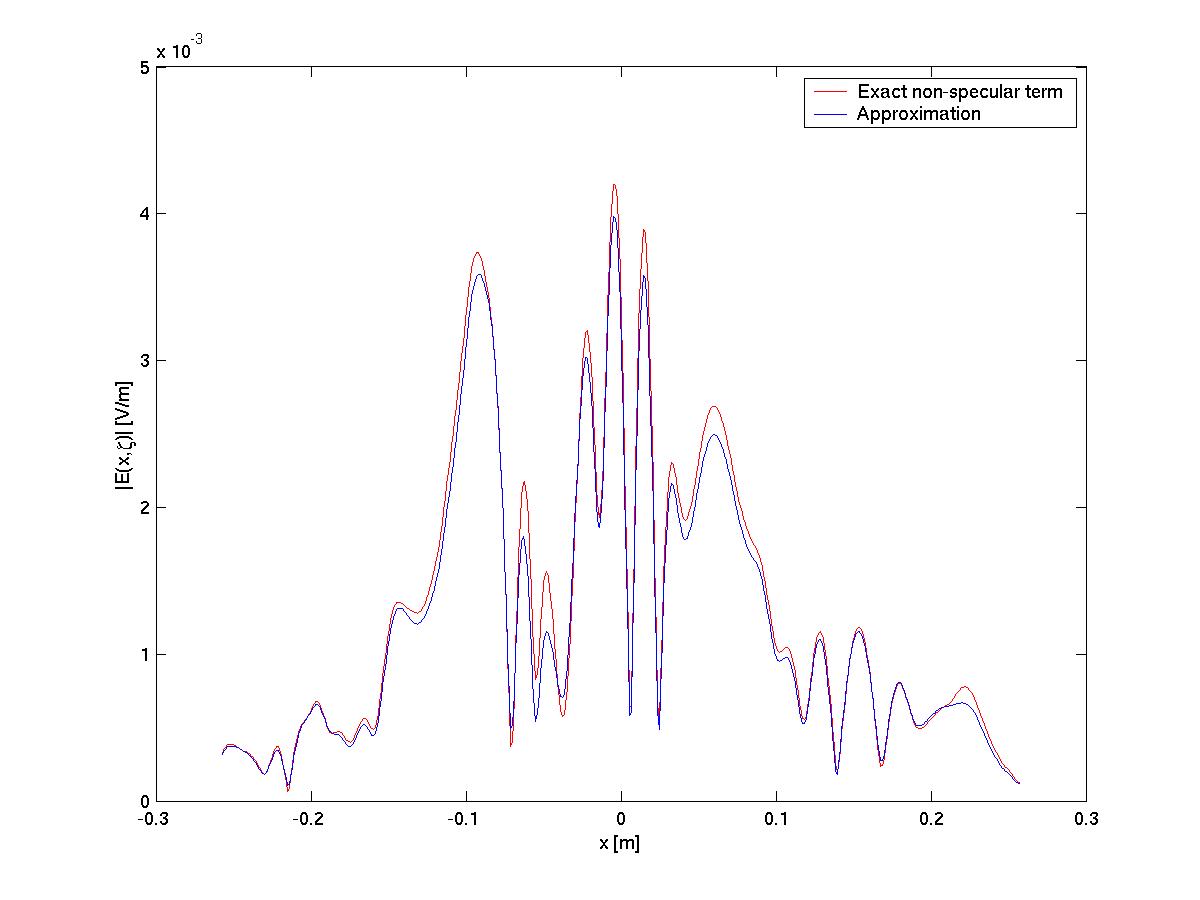} 
\caption{Comparison of exact numerical solution and approximation 
for scattered field on rough profile $z=\zeta$ with varying
impedance $Z=Z_0+Z_r$ for TE polarisation.}
\label{6o6fig1}
\end{figure}

Figure \ref{6o6fig1} compares the surface field term
$-{\Ccal}_0^{-1}{\Ccal}_1{\Ccal}_0^{-1}E_{inc}(\r)$
with the corresponding component of the `exact' 
numerical solution of (\ref{eq:8.10'}). 
Here the angle of incidence is around $5^o$, the ratio of
r.m.s. surface height to wavelength
$<\zeta^2>^{1/2}/\lambda=2/3$, and the ratio of r.m.s. impedance
variation to reference value $Z_0$ is around 1/6.  Agreement is
seen to be very close.

\subsection{Plane boundary with varying impedance\label{flat}}

We now consider the special case of a planar surface $\zeta(x)\equiv
0$ with variable impedance,  
using the operator expansion above.  We should mention here the elegant
method of \cite{dragna} and that of \cite{ostashev} which also
considers ensemble averages and could alternatively be employed. 
To simplify notation we will
denote the operators in this case by $\Azero$ and $\Aone$ so that
equation~(\ref{inteq}) becomes 
\begin{equation}E_{inc}(\r)=({\Acal}_0+{\Aone})E(\r) ,
\label{eq:8.10}\end{equation}  
where $\r$ lies on the surface, ${\Acal}_0$  and ${\Aone}$ are now given by 
\begin{equation}{\Acal}_0(~\cdot~)={1\over
2}(~\cdot~)-\int_{z=0}\left[{\partial G(\r,\r')\over\partial
z}+{ik_0G(\r,\r')\over
Z_0}\right](~\cdot~)dS'\rm\label{eq:8.5}\end{equation} 
and
\begin{equation}{\Aone}(~\cdot~)={ik_0\over
Z_0}\int_{z=0}{Z_r(x')G(\r,\r')\over
Z_0+Z_r(x')}(~\cdot~)dS'.\label{f6.19}\end{equation} 
Eq. (\ref{eq:8.13}) then becomes
\begin{equation}E(\r)\cong
{\Azeroinv}E_{inc}(\r)~-~
{\Azeroinv}\left[{\Aone} {\Azeroinv}E_{inc}(\r)\right].\label{Aflat} 
\end{equation}

The solution to (\ref{eq:8.10}) represents the total field at $z=0$; 
from this the field elsewhere can be obtained 
by writing $E(\r)$ as a superposition of plane waves without recourse
to the integral (\ref{inteq2}).

Suppose for the moment that the impedance is constant, 
$Z=Z_0$, so that $\Aone$ vanishes.
For an incident plane wave, say 
$E_\theta(x,z)=\exp(i k[\sin\theta x - \cos\theta z])$
at an angle $\theta$ with respect to the normal,
the solution is explicitly
\begin{equation}{\Azeroinv}E_\theta(x,0)~=~
\left[1+R(\alpha)\right]\exp(i\alpha x),
\label{6j:6.7}
\end{equation} 
where $\alpha=k\sin\theta$, $\beta=\sqrt{k^2-\alpha^2}$, and
$R$ is the reflection coefficient 
\begin{equation}R(\alpha)={\beta Z_0-k_0\over\beta Z_0+k_0} . 
\label{ampEr}\end{equation}
Thus $\Azeroinv f$ can be found
for {\sl arbitrary} $f(x)$ by expressing $f$ as a
superposition of plane waves and applying (\ref{6j:6.7}). 
If impedance variation $Z(x)=Z_0+Z_r(x)$ is now reintroduced,
then (\ref{eq:8.10}) has formal solution
\begin{equation}E(\r)=({\Acal}_0+{\Aone})^{-1}E_{inc}(\r
)  . \label{eq:8.11}\end{equation}
The first term on the right of equation (\ref{eq:8.13})
is the known specular reflection from a constant impedance surface at $z=0$;
the second models its diffuse modification due to $Z_r$,
i.e. diffraction effects due to impedance variation.

\begin{figure}
\hskip 1.5cm 
\includegraphics[height=7cm]{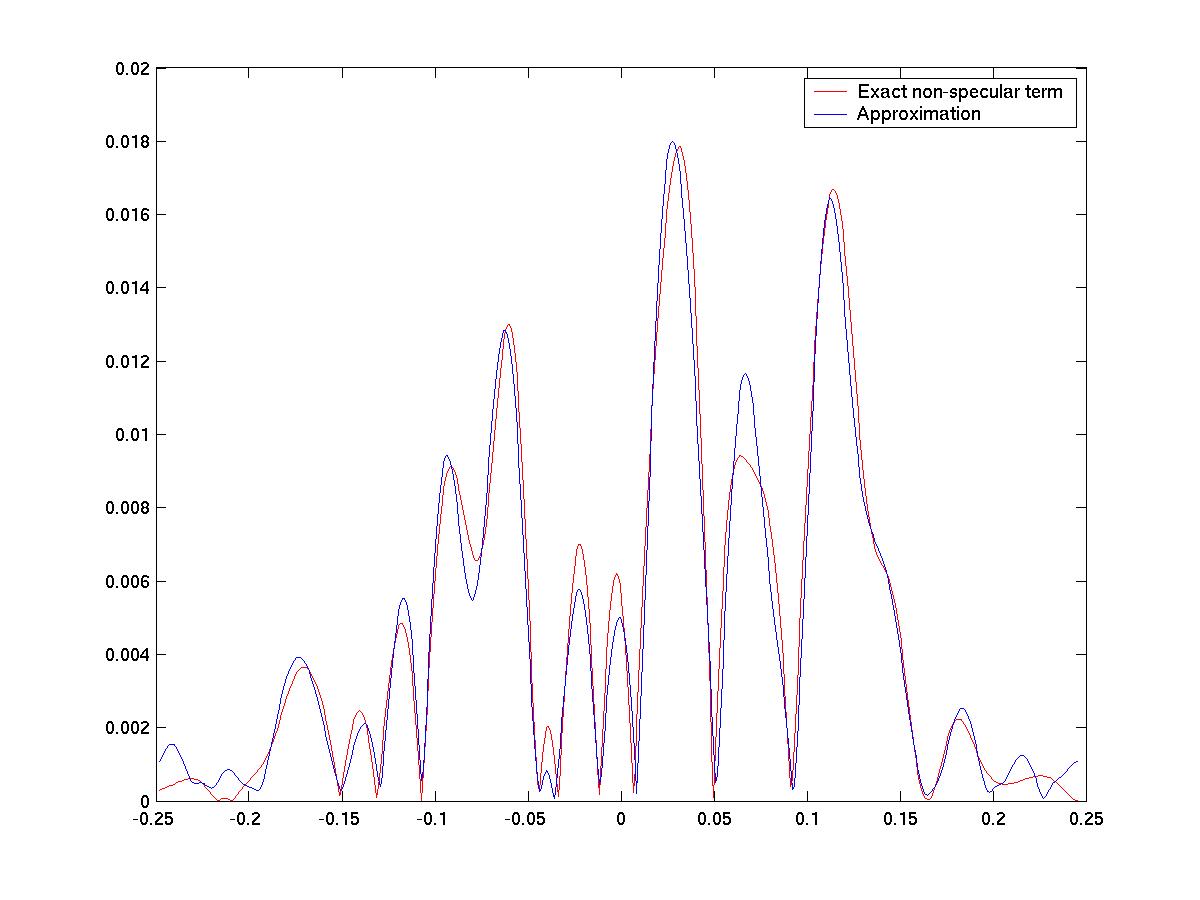} 
\caption{Comparison of exact and approximate solutions for 
non-specular component of scattered field for a flat surface with
varying impedance.}
\label{6o3fig1}
\end{figure}

Specifically, from (\ref{6j:6.7}) and (\ref{f6.19}) we obtain  
\begin{equation}{\Aone}({\Azeroinv}E_{inc}(\r))=
{ik_0(1+R(\alpha))\over Z_0}\int_{z=0}{Z_r(x')G(\r,\r')\over
Z_0+Z_r(x')}e^{i\alpha x'}dS'\rm.\label{eeee6.21}\end{equation} 
As ${\Azeroinv}$ represents reflection by constant
impedance, (\ref{eeee6.21}) 
can be thought of as a secondary `driving field' for the diffuse term
in (\ref{eq:8.13}). 
This field consists of a set of plane waves determined by the Fourier
transform of the integral in (\ref{eeee6.21}).
An example is shown in Figure \ref{6o3fig1} comparing this term in
(\ref{eq:8.13}) with the diffuse part of the exact numerical solution.
$Z_r$ here has an rms value of around
$Z_0/4$. 
(Note that in these simulations the incident field has been tapered to
zero at the edges to minimise spurious edge-effects.)

As the solution of ${\Azeroinv}$ is known the expression
(\ref{eq:8.13}) can be evaluated directly, for one or many 
realisations, and avoids a potentially expensive numerical inversion.

\par

\section{Coherent field\label{coherent}}

In this section we consider statistics of the scattered field (a) when
the average is taken over the ensemble of varying impedance functions and the profile is
deterministic but arbitrary, and (b) the more general case of a
randomly rough surface, taking the average over both impedance and surface profiles.

As the impedance is often known statistically rather than individually, 
evaluation of the mean field is important.
For {\it flat} surfaces the mean field with respect to an ensemble of
impedance realisations obeys an effective impedance
condition, for which an approximation is derived in section (\ref{meanplane}).  
Thus for an incident plane wave the
mean scattered field is specular, but with an `effective reflection
coefficient' depending on incident angle.
For a given {\it rough} surface, the coherent field is no longer specular, and its  
description is therefore more complex.

\subsection{Mean field for plane boundary \label{meanplane}} 

We first consider the coherent field for a flat varying-impedance surface, \S \ref{flat}.
A low-order approximation for the mean field due to scattering
by the randomly varying impedance is easily derived
from the expansion (\ref{eq:8.13}) in this case.
As $Z_r(x)$ is statistically stationary, the coherent field 
is specular and takes the form of a constant effective impedance whose value we seek.
Averaging equation (\ref{Aflat}) gives the mean field at the surface
\begin{equation}<E(\r)>\cong
{\Azeroinv}E_{inc}(\r)-{\Azeroinv}<{\Aone}>
{\Azeroinv}E_{inc}(\r)
\label{x333}\end{equation} 
as the term 
${\Azeroinv}E_{inc}(\r)$ on which $\Aone$ acts is independent of
$Z_r$.
From equation (\ref{f6.19}), 
$<{\Aone}>$  is given by
\begin{equation}<{\Aone}(~\cdot~)>~=~
{ik_0\over Z_0}\int_{z=0} ~\Zmean~ G(\r,\r')(~\cdot~) dS' ~=~
\Zmean~{ik_0\over Z_0}\int_{z=0} ~ G(\r,\r')(~\cdot~) dS'
\label{x334}\end{equation} 
where 
\begin{equation} \Zmean ~=~ \left<\frac{Z_r}{Z_0+Z_r}\right> ~=~ 
1 - Z_0 \left<\frac{1}{Z_0+Z_r}\right> 
\label{7.40}\end{equation} 

Note that this quantity is a one-point average, which does not depend on the 
impedance autocorrelation $\rho(\xi)=<Z_r(x)Z_r(x+\xi)>$. 
This scalar can be found analytically for a wide range of distributions and in any case
numerical averaging is straightforward and rapid for arbitrary statistics.  
For analytical evaluation, $Z_r$ is
most commonly assumed to obey a modified form of complex Gaussian distribution.
If the distribution is {\it exactly} Gaussian then
the probability integral has a pole at $Z_r=-Z_0$, but also
a well-defined Cauchy principal value, and can be obtained
analytically.   However, the singularity at $-Z_0$ corresponds to
vanishing impedance which may be excluded on physical grounds, and the
distribution can be replaced by a Gaussian with cut-off.

Alternatively, to fourth order in the ratio $Z_r/Z_0$, $\Lambda$ can be written
\begin{equation}
\Lambda ~\cong~ \left<{Z_r/Z_0 ~-~ Z_r^2/Z_0^2 ~+~ Z_r^3/Z_0^3
~-~ Z_r^4/Z_0^4 }\right>
\end{equation}
giving a simple high-contrast second order approximation:
\begin{equation} \Zmean ~\cong~ \left<{Z_r^2\over Z_0^2}\right> ~=~ - \sigma^2_I
\label{meanapprox}\end{equation} 
valid for any distribution, or a fourth order approximation for the Gaussian case: 
\begin{equation} \Zmean ~\cong~ -\left<{Z_r^2\over Z_0^2}\right> ~=~ -\sigma^2_I + 3\sigma^4
\label{meanapprox4}\end{equation} 

In any case, when (\ref{x334}) and (\ref{7.40}) are substituted
back into equation (\ref{x333}) it is easily seen
that the mean field is equivalent to a solution of the original problem with
a modified or `effective' constant impedance $\Zf=Z_0+\Zm$, with
$\Zm$ given by
$\Zm =Z_0 \Zmean/(1-\Zmean)$. This immediately gives an effective
reflection coefficient 
\beq
\Rf(\alpha)={\beta \Zf-k_0\over\beta \Zf+k_0}
\eeq
where
\beq
\Zf~=~ \frac{Z_0}{1-\Zmean} .
\label{zeff}
\eeq

\subsection{Mean field for a rough surface \label{rough}}

At a point $\r$ along a given horizontal line above the surface,
the field is related to the surface values via the 
integral (\ref{inteq2}), which is written
\beq
E(\r)= \Cf E 
\label{eq1}
\eeq
where $\Cf$ is the integral operator in (\ref{inteq2}) and the prime is simply to distinguish the operator $\Ccal$ evaluated away from the surface from its value along the surface as occurring in (\ref{inteq}).
If $\Cf$ is split as before into its
constant and varying impedance parts $\Cf_0$ and $\Cf_1$, then,
using (\ref{eq:8.13}), equation (\ref{eq1}) can be written as
\begin{eqnarray}
E &=& \Cf\Ccal^{-1} E_{inc} \nonumber\\
 &=& (\Cf_0+\Cf_1)(\Ccal_0+\Ccal_1)^{-1} E_{inc} \nonumber\\
 &\cong& (\Cf_0+\Cf_1)
(\Ccal_0^{-1}- \Ccal_0^{-1}\Ccal_1\Ccal_0^{-1})E_{inc} \nonumber\\
 &\cong& \left[ \Cf_0 \Ccal_0^{-1}
-\Cf_0 \Ccal_0^{-1}\Ccal_1\Ccal_0^{-1} + \Cf_1 \Ccal_0^{-1})
\right]E_{inc}
\label{eq3}
\end{eqnarray}
where we have neglected a term of higher order in $\Ccal_1$.  
We can now take an
ensemble average of (\ref{eq3}) with respect to impedance variation,
to get the mean modification by impedance variation
of the scattered fields.  
\beq
<E(\r)> ~\cong~\left[ \Cf_0 \Ccal_0^{-1}
~+~ \Cfmean \Ccal_0^{-1} ~-~\Cf_0 \Ccal_0^{-1} \Cmean \Ccal_0^{-1})
\right]E_{inc}
\label{mean}
\eeq
where $\r$ is in the medium and for compactness $\Cmean$ denotes the mean $<\Ccal_1>$ and thus
\beq
\Cmean(~\cdot~) ~=~ {ik_0\over
Z_0}\int_{z=\zeta(x)}\Zmean G(\r,\r')(~\cdot~)dS'
~=~ \Zmean ~\Cone,
\label{ccal2}
\eeq
$\Cfmean$ defined similarly, and $\Zmean$ is given by eq. (\ref{7.40}). 

Expression (\ref{mean}) gives the mean field for an arbitrary irregular surface with
randomly varying impedance, but as $\Cmean$, $\Cfmean$ depend on the
surface profile $\zeta(x)$, numerical evaluation cannot in general be avoided. 
In particular this gives rise to a coherent field spectrum with
effective coefficients depending on the surface profile.  
The approximation (\ref{mean}) is equivalent to the solution $E_e$,
say, for scattering by the surface $\zeta(x)$ but with constant effective
impedance $\Zf$.   This is easily seen by formulating this equivalent problem
in terms of integral operators where it becomes
\beq
E_e(\r) ~=~ (\Cf_0+\Zmean\Cf)(\Ccal_0+\Zmean\Cone)^{-1} E_{inc} ,
\label{E_eff}
\eeq
and then solving as before and comparing terms with (\ref{mean}). 
In terms of the effective field evaluated along the surface which we can denote $E_{se}$,
(\ref{E_eff}) becomes
\beq
E_e(\r) ~=~ (\Cf_0+\Zmean\Cf) ~E_{se}
\label{E_eff2}
\eeq
In other words the effective field $E_e$, i.e. average over impedance realisations, is the solution to the boundary problem given by equations (\ref{inteq}) and (\ref{inteq2}) with varying impedance replaced by effective impedance $\Zf$:
\begin{equation}E(\r) ~=~
~\int_{z=\zeta(x)}\left[{\partial G(\r,\r')\over\partial
n}+{ik_0G(\r,\r')\over
\Zf}\right] E(\r')dS'\rm.\label{inteq3}\end{equation}

\subsection{TM case \label{TM}}
The results above apply to a TE incident field. It is straightforward to derive equivalent results for TM incidence as follows.  Integral equations (\ref{inteq}) and \ref{inteq2}) for $E$ are replaced by 
the following equations for the field $H$: 
\begin{equation}H_{inc}(\r_s) ~=~
\frac{1}{2}H(\r_s)~-
~\int_{z=\zeta(x)}\left[{\partial G(\r_s,\r')\over\partial
n}+{ik_0G(\r_s,\r')
(Z_0+Z_r(x'))}\right] H(\r')dS'\rm.\label{inteqTM}\end{equation} 
where $\r_s$ is an arbitrary surface point $(x,\zeta(x))$, and 
$\r'=(x',\zeta(x'))$. Elsewhere the field can be written as a boundary integral: 
\begin{equation}H(\r) ~=~
~\int_{z=\zeta(x)}\left[{\partial G(\r,\r')\over\partial
n}+{ik_0G(\r,\r')
(Z_0+Z_r(x'))}\right] H(\r')dS'\rm.\label{inteq2TM}\end{equation} 

Following analogous reasoning we obtain
\begin{equation}H(\r)\cong
{\Dzeroinv}H_{inc}(\r)~-~
{\Dzeroinv}\left[{\Done} {\Dzeroinv}H_{inc}(\r)\right].\label{eq:8.13TM} 
\end{equation} 
where now
\begin{equation}{\Dcal}_0(~\cdot~)={1\over
2}(~\cdot~)~-~\int_{z=\zeta(x)}\left[{\partial G(\r,\r')\over\partial
n}+{ik_0 Z_0 G(\r,\r')}\right](~\cdot~)dS'\rm. 
\label{eq:8.5TM}
\end{equation} 
\begin{equation}
{\Dcal}_1(~\cdot~)~=~ ~-~{ik}\int_{z=\zeta(x)}{Z_r(x')G(\r,\r')}(~\cdot~)dS'.
\label{dcal1TM}
\end{equation} 
It is immediately clear, however, that when taking the mean with
respect to impedance variation the term $<{\Dcal}_1>$ vanishes so that
to this order the effective impedance coincides with $Z_0$. Thus
${\Dcal}_1$ has no effect on $<H>$. The effect on the autocorrelation
of $H$ and therefore on mean intensity will be non-zero, but this is
beyond the scope of the present study. 

\subsection{Averaging over rough surfaces \label{mean2}}
Provided the surface profile and impedance are statistically
independent, the above results (\ref{E_eff2}) or (\ref{inteq3}) can be
used to examine the double average $\ll E \gg$ with respect to rough
surface and impedance variation. This may be done using results
existing in various regimes, which we will not reproduce in detail
here.    To illustrate this, 
consider an incident plane wave at angle of $\theta$ for small surface
height $\sigma^2_S <  1$. Perturbation theory to first order in
surface height can be applied along the lines of \cite{watson}.  This
allows the first order (in $\sigma_S$) component to be written as
$\zeta(x) u(\theta,\Zf)$ 
where $u(\theta,\Zf)$ is a known function of incident angle and
effective impedance. 

From this the field everywhere can be expressed in the standard way in
terms of the spectral components via the Fourier transform of $\zeta$,
with the function $u(\theta,\Zf)$ present as a multiplying
factor. Using this we can obtain coherent field and field correlation
within the small surface height regime.    Similarly, the mean field
for low grazing angle incident waves may be obtained by extending
results such as \cite{spivack1} although these require further
development.


\section{Conclusions\label{conclusions}}

Wave scattering by a rough surface with
random spatially varying impedance has been considered. We have sought an
efficient method for calculating the field while allowing convenient
estimation of the effects of impedance variation and
its interaction with the surface profile.

The expressions obtained also provide estimates 
of the mean field with respect to impedance variation.
For rough surfaces these are semi-analytical in the sense that
numerical evaluation of integrals is needed.
(In the case of a flat surface, for which the coherent field is specular,
this takes the form of an effective impedance; this is also
approximately true for a given irregular surface, 
but the behaviour is more complicated because of the 
non-specular nature of the scattered field.)

For simplicity we have restricted attention to 2-dimensional problems,
but the extension to full 3-dimensional scattering is
straightforward.  Similarly,  equivalent results for a TM fields are
easily obtained, and the acoustic case follows immediately.
A further question which is not addressed here is of the
coherent field which results from ensemble of randomly rough surfaces
with varying impedance. A key difficulty is that typically when 
this occurs in practice the roughness and impedance are not
statistically independent.

\vskip 2 true cm
 \subsection{Acknowledgments} 

MS gratefully acknowledges partial support for this work under US
ONR Global NICOP grant N62909-19-1-2128.

\bibliography{references}

\begin{thebibliography}{10}

\bibitem{dragna}
D~Dragna and P~Blanc-Benon.
\newblock Sound propagation over the ground with a random spatially-varying
  surface admittance.
\newblock {\em The Journal of the Acoustical Society of America},
  142(4):2058--2072, 2017.

\bibitem{ostashev}
Vladimir~E Ostashev, D~Keith Wilson, and Sergey~N Vecherin.
\newblock Effect of randomly varying impedance on the interference of the
  direct and ground-reflected waves.
\newblock {\em The Journal of the Acoustical Society of America},
  130(4):1844--1850, 2011.

\bibitem{sarabandi2}
K~Sarabandi and T~Chiu.
\newblock Electromagnetic scattering from slightly rough surfaces with
  inhomogeneous dielectric profiles.
\newblock {\em IEEE Transactions on Antennas and Propagation},
  45(9):1419--1430, 1997.

\bibitem{hatziioannou}
Yannis Hatziioannou.
\newblock Scattering of an electromagnetic wave by a conducting surface.
\newblock {\em Journal of Modern Optics}, 46(1):35--47, 1999.

\bibitem{giovannini}
H~Giovannini, M~Saillard, and A~Sentenac.
\newblock Numerical study of scattering from rough inhomogeneous films.
\newblock {\em JOSA A}, 15(5):1182--1191, 1998.

\bibitem{guerin}
Charles-Antoine Gu{\'e}rin and Anne Sentenac.
\newblock Second-order perturbation theory for scattering from heterogeneous
  rough surfaces.
\newblock {\em JOSA A}, 21(7):1251--1260, 2004.

\bibitem{depine1}
Ricardo~A Depine.
\newblock Backscattering enhancement of light and multiple scattering of
  surface waves at a randomly varying impedance plane.
\newblock {\em JOSA A}, 9(4):609--618, 1992.

\bibitem{brud1}
VL~Brudny and RA~Depine.
\newblock Theoretical study of enhanced backscattering from random surfaces.
\newblock {\em Optik}, 87(4):155--158, 1991.

\bibitem{blumberg1}
DG~Blumberg, V~Freilikher, I~Fuks, Yu~Kaganovskii, AA~Maradudin, and
  M~Rosenbluh.
\newblock Effects of roughness on the retroreflection from dielectric layers.
\newblock {\em Waves in random media}, 12(3):279--292, 2002.

\bibitem{brown}
Gary~S Brown.
\newblock Special issue on low-grazing-angle backscatter from rough surfaces.
\newblock {\em IEEE Transactions on Antennas and Propagation}, 46(1):1--2,
  1998.

\bibitem{hatziioannou2}
Y~Hatziioannou and M~Spivack.
\newblock Electromagnetic scattering by refractive index variations over a
  rough conducting surface.
\newblock {\em Journal of Modern Optics}, 48(7):1151--1160, 2001.

\bibitem{li}
Eric~S Li and Kamal Sarabandi.
\newblock Low grazing incidence millimeter-wave scattering models and
  measurements for various road surfaces.
\newblock {\em IEEE Transactions on Antennas and Propagation}, 47(5):851--861,
  1999.

\bibitem{ogilvy}
JA~Ogilvy and Institute of~Physics~(UK).
\newblock {\em Theory of wave scattering from random rough surfaces}.
\newblock CRC Press, 1991.

\bibitem{voronovich}
AG~Voronovich.
\newblock {\em Wave scattering from rough surfaces}, volume~17.
\newblock Springer Science \& Business Media, 2013.

\bibitem{watson}
John~G Watson and Joseph~B Keller.
\newblock Reflection, scattering, and absorption of acoustic waves by rough
  surfaces.
\newblock {\em The Journal of the Acoustical Society of America},
  74(6):1887--1894, 1983.

\bibitem{basra}
Narinder~Singh Basra.
\newblock {\em Wave Scattering by Rough Surfaces with Varying Impedances.}
\newblock PhD thesis, University of Cambridge, 2003.

\bibitem{desanto}
JA~DeSanto.
\newblock Exact boundary integral equations for scattering of scalar waves from
  infinite rough interfaces.
\newblock {\em Wave Motion}, 47(3):139--145, 2010.

\bibitem{Kapp}
David~A Kapp and Gary~S Brown.
\newblock A new numerical method for rough-surface scattering calculations.
\newblock {\em IEEE Transactions on Antennas and Propagation}, 44(5):711, 1996.

\bibitem{spivack1}
M~Spivack.
\newblock Moments of wave scattering by a rough surface.
\newblock {\em The Journal of the Acoustical Society of America},
  88(5):2361--2366, 1990.

\bibitem{spivack2}
M~Spivack, A~Keen, J~Ogilvy, and C~Sillence.
\newblock Validation of left--right method for scattering by a rough surface.
\newblock {\em Journal of Modern Optics}, 48(6):1021--1033, 2001.

\bibitem{brennan}
Connor Brennan, Peter Cullen, and Marissa Condon.
\newblock A novel iterative solution of the three dimensional electric field
  integral equation.
\newblock {\em IEEE Transactions on Antennas and Propagation},
  52(10):2781--2785, 2004.

\end{thebibliography}

\end{document}